\documentclass[12pt]{article}
\usepackage[]{epsfig}
\usepackage[centertags]{amsmath}
\usepackage{amsfonts}
\usepackage{amssymb}
\usepackage[english]{babel}
\usepackage{amsmath, amsthm, slashed,url}
\usepackage{color}
\begin{document}

\title{\bf A note on degeneracy of excited energy levels in massless Dirac fermions  }
\author{ 
Lucas~Sourrouille
\\
{\normalsize\it  IFLP-CONICET
}\\ {\normalsize\it Diagonal 113 y 64 (1900), La Plata, Buenos Aires, Argentina}
\\
{\footnotesize  lsourrouille@iflp.unlp.edu.ar} } \maketitle

\abstract{We propose a mechanism to construct the eigenvalues and eigenfunctions of the massless Dirac-Weyl equation in the presences of magnetic flux $\Phi$ localized in a restricted region of the plane. Using this mechanism we analyze the degeneracy of the existed energy levels. We find that the zero and first energy level 
has the same $N+1$ degeneracy, where $N$ is the integer part of $\frac{\Phi}{2\pi}$. In addition, and contrary to what is described in the literature regarding graphene, we show that
higher energy levels are $N+m$ degenrate, beign $m$ the level of energy. In other words, this implies an indefinite growth of degenerate states as the energy level grows.

}
 
\vspace{0.3cm}


\vspace{1cm}
\section{Introduction}
The reality of massless Dirac fermions in graphene has been demonstrated by
Novoselov et al.\cite{Novo} and Zhang et al.\cite{Zhang} using quantized magnetic
fields. The discovery of the anomalous (half-integer) quantum Hall effect in
these works was the real beginning of the great interest in graphene.
Also, graphene is an appropriate material to develop electronic
devices.
Recently a series of studies concerning the
interaction of graphene electrons in perpendicular magnetic
fields have been carried out in order to find a way for
confining the charges \cite{Pee}-\cite{KNNL}. In these works the Dirac-Weyl
equation for massless electrons with a Fermi velocity $v_F$ is
considered, where a minimal coupling with the vector potential
describes the interaction with the external field. In general,
some kinds of numerical computation are needed to find the
energy levels of confined states or transmission coefficients for
scattering states.
\\
On the other hand, the energy spectrum of massless Dirac fermions coupled to gauge fields has a very special property, the existence of
zero-energy modes. This is a consequence index theorem \cite{AS}-\cite{AS1}. The existence of the zero-energy states for massless Dirac fermions in inhomogeneous magnetic
fields in two dimensions was demonstrated explicitly by Aharonov and Casher \cite{AC}.
\\
In this note, we study the degeneracy of energy levels of the Dirac-Weyl massless fermions. We begin by review the result of Aharonov and Casher. Then, we propose a mechanisms to obtain, from the zero modes, the existed eigenvalues and eigenstates of the  Dirac-Weyl equation. We analize the integrability of the existed states. We conclude that the zero and first energy level 
has the same $N+1$ degeneracy, where $N$ is the integer part of $\frac{\Phi}{2\pi}$, whereas the rest of de levels are $N+m$ degenrate, beign $m$ the level of energy. This is a novel result, since contrary to what is described in the literature regarding graphene, this implies an indefinite growth of degenerate states as the energy level grows.

\section{The framework and the Aharonov-Casher theorem}

Let us start by considering a $(2+1)$-dimensional Dirac-Weyl model whose Hamiltonian is described by
\begin{equation}
H= \sigma^j p_j = (\sigma^1 p_1 +  \sigma^2 p_2)\;,
\label{}
\end{equation}
Here, the $\sigma^j$ $(j =1,2)$ 
are 2$\times$2 Pauli matrices, i.e.
\begin{eqnarray}
\sigma^1 =\left( \begin{array}{cc}
0 & 1 \\
1 & 0 \end{array} \right)
\,,
\;\;\;\;\;\
\sigma^2 =\left( \begin{array}{cc}
0 & -i \\
i & 0 \end{array} \right)
\end{eqnarray}
and $p_j =-i\partial_j$ is the two-dimensional momentum operator. The massless Dirac-Weyl equation in $(2+1)$ dimensions is
\begin{equation}
\sigma^j p_j \Phi (x, y, t) = i\partial_t \Phi (x, y, t)
\label{eq1}
\end{equation}
Here, $\Phi (x, y, t)$ is the two-component spinor
\begin{equation}
\Phi=(\phi_a,\phi_b)^T
\label{}
\end{equation}
where $\phi_a$ and $\phi_b$ represent the envelope functions associated with the probability amplitudes. Since, we are interested 
in stationary states, it is natural to propose a 
solution of the form
\begin{eqnarray}
\Phi (x, y, t) = e^{-iEt} \Psi (x, y)\;,
\label{}
\end{eqnarray}
then, the time-independent Dirac-Weyl equation is
\begin{equation}
\sigma^j p_j \Psi (x, y) = E \Psi (x, y)
\label{1dw}
\end{equation}
In the presences of a perpendicular magnetic field to the $(x, y)$-plane, we replace the momentum operator $p_j$ by the covariant 
derivative, defined as $D_{j}= -i\partial_{j} +A_{j}$ $(j =1,2)$, where $A_{j}$ are components of the vector potential,
\begin{equation}
B=\partial_x A_y -\partial_y A_x
\label{mag}
\end{equation}
Thus, the equation (\ref{1dw}) becomes, 
\begin{equation}
\sigma^j D_j \Psi (x, y) = E \Psi (x, y)
\label{2dw}
\end{equation}
We can develop this equation to get,
\begin{equation}
\left( \begin{array}{cc}
0 & D_1 -iD_2 \\
D_1 +iD_2 & 0 \end{array} \right) \left( \begin{array}{c}
\psi_a\\
\psi_b \end{array} \right) = E \left( \begin{array}{c}
\psi_a\\
\psi_b \end{array} \right) 
\label{3dw}
\end{equation}
where $\psi_a$ and $\psi_b$ are the components of the spinor $\Psi$ (i.e. $\Psi=(\psi_a,\psi_b)^T$). From this equation we can
write the two coupled equations for the components $\psi_a$ and $\psi_b$
\begin{eqnarray}
D_1 \psi_b -iD_2 \psi_b = E \psi_a
\label{eqm1}
\end{eqnarray}
\begin{eqnarray}
D_1 \psi_a +iD_2 \psi_a = E \psi_b
\label{eqm2}
\end{eqnarray}
For the energy zero modes we can construc the solutions following the work done by 
Aharonov and Casher \cite{AC},
where the magnetic flux is
localized in a restricted region
For this purpose we assume that the vector potential is divergenceless, that is 
\begin{eqnarray}
\partial_xA_x +
\partial_yA_y =0
\label{12}
\end{eqnarray}
Note that Eqs. (\ref{mag}) and (\ref{12}) which define the
magnetic field and the gauge condition imply that
${\bf A}$ is a two-dimensional curl whose potential
satisfies the Laplace equation with $B(x,y)$ as a source:
\begin{eqnarray}
A_x = -\partial_y \lambda
\,,
\;\;\;\;\;\
A_y = \partial_x \lambda
\label{gau}
\end{eqnarray}
and due to the equation (\ref{mag}),
\begin{eqnarray}
B = \partial_x^2 \lambda + \partial_y^2 \lambda
\label{div}
\end{eqnarray}
Using (\ref{gau}), we can rewrite equations (\ref{eqm1}) and (\ref{eqm2}) as follow
\begin{eqnarray}
(-i\partial_x -\partial_y) \psi_b + (-i\partial_x \lambda -\partial_y \lambda) \psi_b  = E \psi_a
\label{e1}
\\[3mm]
(-i\partial_x +\partial_y) \psi_a - (-i\partial_x \lambda +\partial_y \lambda) \psi_a  = E \psi_b
\label{e2}
\end{eqnarray}
Then, it is not difficult to find the solutions of the equations (\ref{e1}) and (\ref{e2}) for the energy zero case. 
Indeed, substituting in equation (\ref{e1})  
\begin{eqnarray}
\psi_b = f_b e^{-\lambda}
\end{eqnarray}
and setting $E=0$, we obtain,
\begin{eqnarray}
[-i\partial_x -\partial_y] f_b=0
\label{fb}
\end{eqnarray}
In similar way, if we propose 
\begin{eqnarray}
\psi_a = f_a e^{\lambda}
\end{eqnarray}
equation is reduced to 
\begin{eqnarray}
[-i\partial_x +\partial_y] f_a=0
\label{}
\end{eqnarray}
Thus, $f_a$ and $f_b$ are analytic and complex conjugated analytic entire functions of $z = ix + y$, respectively. 
\\
The equation (\ref{div}) has the following solution 
\begin{eqnarray}
\lambda ({\bf r}) = \int  d {\bf r}'G({\bf r}, {\bf r}') B({\bf r}')
\end{eqnarray}
where 
\begin{eqnarray}
G({\bf r}, {\bf r}')= \frac{1}{2\pi}\ln \Big(\frac{|{\bf r} - {\bf r}'|}{r_0}\Big)
\end{eqnarray}
is the Green function of the Laplace operator in two dimensions and $r_0$ is an arbitrary constant. 
According to Ref.\cite{AC} the magnetic flux $\Phi$ is localized in a restricted region so that for $r \to \infty$
\begin{eqnarray}
\lambda ({\bf r}) = \frac{\Phi}{2\pi} \ln \Big(\frac{r}{r_0}\Big)\label{23}
\end{eqnarray}
and 
\begin{eqnarray}
\psi_{a,b} = f_{a,b} \Big(\frac{r}{r_0}\Big)^{\frac{\gamma \Phi}{2\pi}}
\label{sol}
\end{eqnarray}
where $\gamma =1$ and $-1$ for $\psi_a$ and $\psi_b$ respectively. Since the entire function $f(z)$ cannot go to zero in all 
directions at infinity, $\psi_{a,b}$ can be normalizable only assuming that $\gamma \Phi < 0$, that is, zero-energy solutions can 
exist only for one spin direction, depending on the sign of the total magnetic flux.
\\
Now, consider the case  $\Phi > 0$, then in view of (\ref{sol}) we have $\psi_a =0$ and 
\begin{eqnarray}
\psi_{b} =  f_{b} e^{-\lambda} \simeq f_{b} \Big(\frac{r}{r_0}\Big)^{\frac{- \Phi}{2\pi}}
\label{25}
\end{eqnarray}
The function $f_b$ is dictated by  (\ref{fb}) and it is not difficult to check that the solutions are polynomials of the form
\begin{eqnarray}
f_{b} = a_j z^j
\end{eqnarray}
where $a_i$ are real numbers.
However, one can easily see from Eq.(\ref{25}) that the solution is integrable with the square only assuming that $j \leq N$, 
where $N$ is the integer part of $\frac{\Phi}{2\pi}$. For the case  $\Phi < 0$ we have
\begin{eqnarray}
\psi_{a} = f_{a} e^{\lambda} \simeq f_{a} \Big(\frac{r}{r_0}\Big)^{\frac{ \Phi}{2\pi}}\,,
\;\;\;\;\;\ \psi_b =0
\label{bb}
\end{eqnarray}
where 
\begin{eqnarray}
f_{a} = \tilde{a}_j (z^\dagger)^j
\end{eqnarray}
and the subscript $j$ must satisfies
\begin{eqnarray}
j \leq N
\label{29}
\end{eqnarray}
Thus, the number of the independent states with zero energy for one spin projection is equal to $N+1$, and there are no 
such solutions for another spin projection. 

\section{Construction of the excited energy states and integrability analysis}
\label{4v}
Let us now concentrate on the construction of eigenstates corresponding to eigenvalues different form zero. Here we develop a mechanism which is valid when the magnetic field is uniform. That is, we suppose the existence of  magnetic field $B$ in a restricted region $A$, so that the magnetic flux is 
\begin{eqnarray}
\Phi = BA
\end{eqnarray}
To proceed we denote
\begin{eqnarray}
D_- = D_1 -iD_2 
\nonumber \\[3mm]
D_+ = D_1 +iD_2 
\label{30}
\end{eqnarray}
So that the equations (\ref{eqm1}) and (\ref{eqm2}) may be rewritten as 
\begin{eqnarray}
D_- \psi_b = E \psi_a
\nonumber \\[3mm]
D_+ \psi_a = E \psi_b
\label{eqm22}
\end{eqnarray}
For simplesity we assume that $\Phi < 0$.
Then, we can show that 
\begin{eqnarray}
\zeta =\left(\begin{array}{c} \hat{E}^{-1}D_- \psi_a \\
\psi_a  \end{array} \right)
\label{32}
\end{eqnarray}
is an eigenstate of the equation (\ref{eqm22}),
where $\hat{E}$ is a real number to be determined. 
In order, to show this we must use the commtator betwen $D_+$ and $D_-$
\begin{eqnarray}
[D_+, D_-] = -2B
\label{33}
\end{eqnarray}
Certainly, we can introduce (\ref{32}) into (\ref{eqm22}), so that $D_-$ acts on the second component of (\ref{32}) and $D_+$ acts on the first. Thus, we have 
\begin{eqnarray}
&&D_- \psi_a = \hat{E} (\hat{E}^{-1}D_-\psi_a)
\nonumber \\[3mm]
&&D_+ D_-(\hat{E}^{-1}\psi_a) = (-2B + D_- D_+) \hat{E}^{-1}\psi_a
\label{}
\end{eqnarray}
Since $D_+ \psi_a = E \psi_b$ and $D_- \psi_b = E \psi_a$, the last equation reduce to
\begin{eqnarray}
D_+ D_-(\hat{E}^{-1}\psi_a) = (E^2-2B)\hat{E}^{-1} \psi_a = \hat{E}^{-1}\hat{E}^2 \psi_a = \hat{E} \psi_a
\end{eqnarray}
where $\hat{E}^2 = E^2-2B$. Therefore, (\ref{32}) is an eigenfunction of the Hamiltonian 

\begin{equation}
H=
\left( \begin{array}{cc}
0 & D_- \\
D_+ & 0 \end{array} \right)  
\label{}
\end{equation}
with eigenvalue
\begin{equation}
\hat{E} = \mp\sqrt{E^2 -2B}
\label{ee}
\end{equation}
Here it is interested to note that we can obtain, from (\ref{ee}), the total spectrum of energy. In fact, since, the ground eigenvalue of energy is zero, we can calculate the firts existed energy level, i.e.
\begin{equation}
E_1 = \mp\sqrt{ -2B}
\label{eee}
\end{equation}
Then, the second level of energy is calculated from $E_1$,
\begin{equation}
E_2 = \mp\sqrt{ -4B} = \mp2\sqrt{ -B}
\label{eeee}
\end{equation}
and so on.
\\
Since, we are assuming that $\Phi < 0$ the zero energy solution is  
\begin{eqnarray}
\zeta_0 =\left(\begin{array}{c} \tilde{a}_j (z^\dagger)^j e^{\lambda} \\
0 \end{array} \right)
\label{}
\end{eqnarray}
We can obtain the first existed state of energy by using formula (\ref{32}), so that 
\begin{eqnarray}
\zeta_1 =\left(\begin{array}{c}\hat{E}^{-1} D_-(\tilde{a}_j (z^\dagger)^j e^{\lambda}) \\
\tilde{a}_j (z^\dagger)^j e^{\lambda} \end{array} \right)
\label{et}
\end{eqnarray}
which is associated to eigenvalue (\ref{eee}). In order to analyze the degeneracy of this state, we develop the first componet of the spinor (\ref{et}), that is, 
\begin{eqnarray}
\hat{E}^{-1} D_-(\tilde{a}_j (z^\dagger)^j e^{\lambda})
\end{eqnarray}
Here, $\hat{E}^{-1} = {E_1}^{-1} = {(\mp\sqrt{ -2B})}^{-1}$, whereas $D_-$ can be developed by using the definition (\ref{30}) and formula (\ref{e1}) 
\begin{eqnarray}
D_-= (-i\partial_x -\partial_y) + (-i\partial_x \lambda- \partial_y \lambda)
\end{eqnarray}
Hence,
\begin{eqnarray}
\hat{E}^{-1} D_-(\tilde{a}_j (z^\dagger)^j e^{\lambda}) =
{(\mp\sqrt{ -2B})}^{-1} 2\tilde{a}_j [(z^\dagger)^{j} \hat\nabla e^\lambda -j(z^\dagger)^{j-1}e^\lambda]
\end{eqnarray}
which leads us to rewrite (\ref{et}) as
\begin{eqnarray}
\zeta_1 =\left(\begin{array}{c}
{(\mp\sqrt{ -2B})}^{-1} 2\tilde{a}_j [(z^\dagger)^{j} \hat\nabla e^\lambda -j(z^\dagger)^{j-1}e^\lambda]
\\
\tilde{a}_j (z^\dagger)^j e^{\lambda} \end{array} \right)
\label{45}
\end{eqnarray}
Here, $\hat\nabla  = (-i\partial_x - \partial_y )$, so that  $\hat\nabla e^\lambda = (-i\partial_x \lambda- \partial_y \lambda)e^\lambda$.  In view of formula (\ref{23}), we can rewrite (\ref{45}), for $r \to \infty$, as
\begin{eqnarray}
\zeta_1 =\left(\begin{array}{c}
{(\mp\sqrt{ -2B})}^{-1} 2\tilde{a}_j [(z^\dagger)^{j} \hat\nabla \Big[\Big(\frac{r}{r_0}\Big)^{\frac{ \Phi}{2\pi}} \Big]-j(z^\dagger)^{j-1} \Big(\frac{r}{r_0}\Big)^{\frac{ \Phi}{2\pi}}  ]
\\
\tilde{a}_j (z^\dagger)^j \Big(\frac{r}{r_0}\Big)^{\frac{ \Phi}{2\pi}} \end{array} \right)
\label{46}
\end{eqnarray}
For simplesity, we can take $r_0= 1$. Then, we can evaluate the term $\hat\nabla \Big[\Big(\frac{r}{r_0}\Big)^{\frac{ \Phi}{2\pi}} \Big]$. In terms of $z$ and $z^\dagger$ we arrive to the following formula
\begin{eqnarray}
\hat\nabla \Big[\Big(\frac{r}{r_0}\Big)^{\frac{ \Phi}{2\pi}} \Big] = \frac{-\Phi}{2\pi} z (z^\dagger z)^{\frac{\Phi}{4\pi} -1}
\end{eqnarray}
Thus, (\ref{46}) reads as 
\begin{eqnarray}
\zeta_1 =\left(\begin{array}{c}
{(\mp\sqrt{ -2B})}^{-1} 2\tilde{a}_j [{\frac{ -\Phi}{2\pi}} (z^\dagger)^{j}  z |z|^{{\frac{- |\Phi|}{2\pi}}-2} -j(z^\dagger)^{j-1} |z|^{\frac{ -|\Phi|}{2\pi}}  ]
\\
\tilde{a}_j (z^\dagger)^j |z|^{\frac{- |\Phi|}{2\pi}} \end{array} \right)
\label{48}
\end{eqnarray}
where $|z|^2 = z^\dagger z = x^2 + y^2$. Also, in view that $\Phi < 0$ we have replace $\Phi$ by $-|\Phi|$. As we see in (\ref{29}), to the second component of the spinor (\ref{48}) be integrable the index $j$ must satisfies the inequality $j \leq N$, which ensure that the number of independent states is $N+1$. A similar analysis can be doing for the first component of the spinor (\ref{48}). In this case it is easy to see that the integrability takes place when $j \leq N +1$. This indicates that we have one more integrable state. However, since (\ref{48}) is a spinor composed of two components, we conclude that the degeneracy is the same as the zero modes, that is $N+1$.  
\\
We can construct the sencond existed states. To proceed we use again the formula  (\ref{32}), where, now, 
\begin{eqnarray}
\psi_a =({\mp\sqrt{ -2B})}^{-1} 2\tilde{a}_j [(z^\dagger)^{j} \hat\nabla e^\lambda -j(z^\dagger)^{j-1}e^\lambda]
\end{eqnarray}
\begin{eqnarray}
\zeta_2 =\left(\begin{array}{c}E_2^{-1} D_-(\psi_a) \\
\psi_a \end{array} \right)
\label{50}
\end{eqnarray}
\begin{eqnarray}
D_-(\psi_a) = \Big[(-i \partial_x -\partial_y) + (-i \partial_x \lambda-\partial_y \lambda)\Big]\Big[({\mp\sqrt{ -2B})}^{-1} 2\tilde{a}_j [(z^\dagger)^{j} \hat\nabla e^\lambda -j(z^\dagger)^{j-1}e^\lambda]\Big]
\end{eqnarray}
To evaluate this expression we use the following results,
\begin{eqnarray}
&&(-i \partial_x -\partial_y)(z^\dagger)^{j-1} = -2(j-1)(z^\dagger)^{j-2}
\nonumber \\[3mm]
&&(-i \partial_x -\partial_y)(z^\dagger)^{j} = -2 j(z^\dagger)^{j-1}
\nonumber \\[3mm]
&&(-i \partial_x -\partial_y)(-i \partial_x \lambda -\partial_y \lambda) = -\partial_x^2 \lambda + 2i \partial_x\partial_y \lambda + \partial_y^2 \lambda
\end{eqnarray}
Then,
\begin{eqnarray}
D_-(\psi_a) = ({\mp\sqrt{ -2B})}^{-1} 2\tilde{a}_j \Big[2(z^\dagger)^{j} (-i \partial_x -\partial_y)^2 e^\lambda -4 j(z^\dagger)^{j-1} \hat\nabla e^\lambda 
\nonumber \\
(z^\dagger)^{j}(-\partial_x^2\lambda + 2i \partial_x\partial_y \lambda + \partial_y^2 \lambda)e^\lambda +2j(j-1)(z^\dagger)^{j-2} e^\lambda
\Big]
\end{eqnarray}
We can develop the term $(-\partial_x^2\lambda + 2i \partial_x\partial_y \lambda + \partial_y^2 \lambda)e^\lambda$
\begin{eqnarray}
(-\partial_x^2\lambda + 2i \partial_x\partial_y \lambda + \partial_y^2 \lambda)e^\lambda =\hat\nabla^2 e^\lambda -  (\hat\nabla\lambda)^2 e^\lambda 
\end{eqnarray}
or equivalently
\begin{eqnarray}
(\hat\nabla^2 \lambda ) e^\lambda = \hat\nabla^2 e^\lambda -  (\hat\nabla\lambda)^2 e^\lambda = \hat\nabla^2 e^\lambda - 4 \hat\nabla e^{\frac{\lambda}{2}}\hat\nabla e^{\frac{\lambda}{2}}
\end{eqnarray}
Therefore, 
\begin{eqnarray}
D_-(\psi_a) = ({\mp\sqrt{ -2B})}^{-1} 2\tilde{a}_j \Big[3(z^\dagger)^{j} \hat\nabla^2 e^\lambda -4 j(z^\dagger)^{j-1} \hat\nabla e^\lambda 
\nonumber \\-
(z^\dagger)^{j} 4 \hat\nabla e^{\frac{\lambda}{2}}\hat\nabla e^{\frac{\lambda}{2}}
+2j(j-1)(z^\dagger)^{j-2} e^\lambda
\Big]
\label{56}
\end{eqnarray}
\begin{eqnarray}
\zeta_2 =\left(\begin{array}{c}(\mp2\sqrt{ -B}
)^{-1} D_-(\psi_a) \\
({\mp\sqrt{ -2B})}^{-1} 2\tilde{a}_j [(z^\dagger)^{j} \hat\nabla e^\lambda -j(z^\dagger)^{j-1}e^\lambda] \end{array} \right)
\label{57}
\end{eqnarray}
As we did above, for $\zeta_1$,  we can analyze the spinor (\ref{57}) for $r \to \infty$. Then, in view of (\ref{23}), we can rewrtite (\ref{56}) as 

\begin{eqnarray}
D_-(\psi_a) = ({\mp\sqrt{ -2B})}^{-1} 2\tilde{a}_j \Big[\frac{3\Phi}{\pi}(z^\dagger)^{j} z^2 |z|^{\frac{-|\Phi|}{2\pi} -4}
+ \frac{2\Phi}{\pi}  j(z^\dagger)^{j-1} z |z|^{\frac{-|\Phi|}{2\pi} -2}
\nonumber \\
- (z^\dagger)^{j} \frac{\Phi^2}{4\pi^2}z^2 |z|^{\frac{- |\Phi|}{\pi} -4}
+2j(j-1)(z^\dagger)^{j-2} |z|^{\frac{- |\Phi|}{2\pi}}
\Big]
\label{58}
\end{eqnarray}
An analyze of this expression shows that the integrability takes place when $j \leq N +2$. 
This means one more state than the first component of (\ref{48}) or equivalently the second componet of (\ref{57}). Therefore, the second level of energy has $N+2$ degenerate states, wich implies one more degenerate state than those of the zero and first energy levels. 
We can repeat this procedure to calculate the eigenstates asociated to higer energy levels. For instance the spinor asociated to the third level of energy is 
\begin{eqnarray}
\zeta_3 =\left(\begin{array}{c}E_3^{-1} D_-(\psi_a) \\
\psi_a \end{array} \right)
\label{59}
\end{eqnarray}
where, now, $\psi_a$ is dictated for the formula (\ref{58}), i.e. 
\begin{eqnarray}
\psi_a = ({\mp\sqrt{ -2B})}^{-1} 2\tilde{a}_j \Big[\frac{3\Phi}{\pi}(z^\dagger)^{j} z^2 |z|^{\frac{-|\Phi|}{2\pi} -4}
+ \frac{2\Phi}{\pi}  j(z^\dagger)^{j-1} z |z|^{\frac{-|\Phi|}{2\pi} -2}
\nonumber \\
- (z^\dagger)^{j} \frac{\Phi^2}{4\pi^2}z^2 |z|^{\frac{- |\Phi|}{\pi} -4}
+2j(j-1)(z^\dagger)^{j-2} |z|^{\frac{- |\Phi|}{2\pi}}
\Big]
\label{60}
\end{eqnarray}
and $E_3 = \pm \sqrt{-6B}$. 
Again, we if want to analyze the integrability of the spinor (\ref{59}), we should apply the operator $D_-$ on the expression (\ref{60}). To proceed we consider the following identities,
\begin{eqnarray}
(-i\partial_x -\partial_y ) |z| = \frac{1}{2}(z^\dagger z)^{-\frac{1}{2}} z = \frac{1}{2}|z|^{-1} z 
\label{61}
\end{eqnarray}
\begin{eqnarray}
(-i\partial_x \lambda -\partial_y \lambda) = -\frac{\Phi}{2\pi}z |z|^{-2} 
\label{62}
\end{eqnarray}
where for the last expression, we have used the formula (\ref{23}) and the fact that $|z|= r$.
The term $\frac{1}{2}|z|^{-1} z$ of equation (\ref{61}) remains constant as $r \to \infty$. This implies that the integrability of $D_-\psi_a$ is only affected by the operator $(-i\partial_x -\partial_y )$, when it acts upon $(z^\dagger)^{l}$, being $l$ an arbitrary integer number. On the other hand the formula (\ref{62}) indicates that the product  
$(-i\partial_x \lambda -\partial_y \lambda)\psi_a$ is integrable when $j \leq N +3$, which implies one more state than the expression (\ref{58}). Since $(-i\partial_x -\partial_y )(z^\dagger)^{l} = -2l (z^\dagger)^{l-1}$, we conclude that $D_-\psi_a$ is integrable for $j \leq N +3$. Therefore, the third level of the energy has $N+3$ degenerate states, which implies one more degenerate state than those of the second energy level. We can continue with this procedure and construct the states $\zeta_4$ asociated to the fourth energy level. Then, using the formulas  (\ref{61}), (\ref{62}) and the same above arguments we can conclude that the fourth energy level is $N+4$ degenerate states. Finally, we can repeat this procedure for  arbitrary number of times, showing that degeneracy goes to infinity when the energy level grows indefinitely. To conclued we note that a similar procedure may be generalized for the case in which $\Phi >0$. In such a case it is not difficult to show that  
\begin{eqnarray}
\zeta =\left(\begin{array}{c}  \psi_b \\
\hat{E}^{-1}D_+\psi_b  \end{array} \right)
\label{63}
\end{eqnarray}
is an eigenstate of the equation (\ref{eqm22}). Thus, if we introduce (\ref{63}) into (\ref{eqm22}) and use (\ref{33}) we have  
\begin{eqnarray}
\hat{E}^{-1}D_-D_+ \psi_b = \hat{E}^{-1} \Big( 2 B + D_+D_-\Big)\psi_b 
\label{64}
\end{eqnarray}
Then, using $ D_-\psi_b = E\psi_a$ and $D_+\psi_a = E\psi_b$, equation (\ref{64}) reads, 
\begin{eqnarray}
\hat{E}^{-1}D_-D_+ \psi_b =
\hat{E}^{-1}( 2B+E^2) \psi_b 
\end{eqnarray}
if we define $\hat{E}^{2} = 2B + E^2$ the last equation is rewritten as 
\begin{eqnarray}
D_-(\hat{E}^{-1}D_+ \psi_b) =
\hat{E} \psi_b 
\end{eqnarray}
On the other hand we can reobtain the second equation of (\ref{eqm22}), 
\begin{eqnarray}
D_+ \psi_b = \hat{E}(\hat{E}^{-1}D_+ \psi_b) 
\end{eqnarray}
In similar way to the case $\Phi <0$ we can obtain the existed eigenvalues from the formula
\begin{eqnarray}
\hat{E} = \pm \sqrt{ 2B + E^2}
\end{eqnarray}
Hence, the first existed energy level is $\hat{E} = \pm \sqrt{ 2B }$, the second  
$\hat{E} = \pm \sqrt{ 4B }$ and so on. 
\\
Thus, we can obtain all eigenfunctions and then analyze the integrability of the these eigenfunctions.

\section{Discussion and Conclusion}

It is interesting to think in the following situation as posible experiment: we can imagine a magnetic flux such that the integer part of $\frac{\Phi}{2\pi}$ is $1$, which implies that the degeneracy of the first two levels ($E_0$ and $E_1$) are $2$, for the second level we have $3$ independent states with energy $E_2= \mp 2 \sqrt{-B}$ (we consider $B<0$ as formula $(40)$ of the paper) and so on. Then we can imagine 3 fermions with energy $E_2$. We can measure the 3 particles with energy $E_2$ (one in each independent state). The system of these 3 particles may be encreased in energy, for instance we can increase the energy from second level to third level and measure the 3 fermions with energy $E_3$. However we can not measure the 3 fermions with energy $E_0$ or $E_1$ since there are 2 independent states associated to these energies. This implies that exist a minimum bound of energy, below which it is not posible measure, at the same time, the 3 fermions in eigenfunctions of the energy.

On the other hand it intersting to note that the filling factor which plays a crucial role in the physics
of the QHE and particularly in QHE in graphene, is given by the following formula  
\begin{eqnarray}
\nu = \frac{\rho\phi_0}{B}
\end{eqnarray}
where $\rho$ is the 2D density of electrons and $\frac{B}{\phi_0}$ is the degeneracy per unit area.
\\
In a classical two dimensional semiconductor the filling factor is measured in relation to the empty lowest Landau level. However, in graphene, since a half of the lowest Landau level is placed in the valence band, the filling factor is not counted with respect to the empty LLL, but with respect to the first empty sublevel of the LLL placed in the conduction band. Thus, such filling in graphene, unlike in a traditional semiconductor is accompanied by two completely filled sublevels of the the LLL located in the valence band \cite{22}-\cite{26}.The relation between graphene and typical semiconductor filling factor can be expressed in a form 
$\nu = \nu_s -2$,
where $\nu_{s}$ is the filling factor of the semiconductor. 
\\
In view of the our results the measure of the filling factor should contrast with this result, since the degeneration increases as energy levels increase.
Thus, it would be intersting measure the filling factor in order to verify the results of the manuscript

Finally, we would like to  remark that the spectrum of the energy found in formula (\ref{ee}) may be derived in a simple way without need to resolve the eigenvalues problem (see section 2.2 of reference \cite{ult}). This simple derivation sugest that the spectrum depend on the magentic field but not on the magnetic flux. In other words, the spectrum does not depend on the area A. However, the eigenfunctions that we found by using formula (\ref{32}) clearly depend, for $r\to \infty$, on the magnetic flux and therefore on the size of the area A. This arises as a consequence of the formula (\ref{23}) which allows us to obtain an expression of the eigenfunctions at $r\to \infty$, and then analyze its integrability to conclue the degeneracy grows as the energy levels grow. In addition, it can be seen that the degeneracy grows as the magnetic flux grows, since as we show the degeneracy, for energy levels higher than the first, are $N+m$, being $N$ the integer part of $\frac{\Phi}{2\pi}$.

In summary, we have developed a formalism to 
construc the eigenfunctions and the eigenvalues of the Dirac-Weyl equation in the presences of magnetic flux $\Phi$ localized in a restricted region of the plane. 
We have shown that for negative magnetic field configurations the eigenstates may be 
constructed from the zero mode with positive chirality whereas for positive magnetic field the eigenstates are 
constructed form the zero mode with negative chirality. This is a consequence of the Aharonov-Casher theorem, which establishes 
that for a negative magnetic flux the zero-energy solutions can exist only for positive spin direction, whereas if the 
magnetic flux is positive the zero-energy solutions can exist only for negative spin direction.
Finally, using this fomalism, we showed that the zero and first energy levels has the same $N+1$ degeneracy and the higher levels of energy are $N+m$ degenerate, being this a novel result, since contrary to what is described in the literature regarding graphene, this implies an indefinite growth of degenerate states as the energy level grows.

\vspace{0.6cm}

{\bf Acknowledgements}
\\
This work is supported by CONICET.

\end{document}